\documentclass[12pt]{article}
\usepackage{latexsym,amsfonts,amsmath,amsthm,amssymb}

\title{Chameleon effect, the range of values hypothesis and
reproducing the EPR-Bohm correlationss}
\author{Luigi Accardi\\
Centro Vito Volterra, University of Rome Tor Vergata,\\ 00133 Rome, Italy\\
Andrei Khrennikov\\
International Center for Mathematical Modeling\\ in Physics,
Engineering and Cognitive science\\ MSI, V\"axj\"o University,
S-35195, Sweden}

\begin{document}

\maketitle

\begin{abstract} 
We present a detailed analysis of assumptions that  J.
Bell used to show that local realism contradicts QM. We find that
Bell's viewpoint on realism is nonphysical, because it implicitly
assume that observed  physical variables coincides with  ontic
variables (i.e., these variables before measurement). The real
physical process of measurement is a process of dynamical
interaction between a system and a measurement device. Therefore one
should check the adequacy of QM not to ``Bell's realism,'' but to
adaptive realism (chameleon realism). Dropping Bell's assumption we
are able to construct a natural representation of the EPR-Bohm
correlations in the local (adaptive) realistic approach.
\end{abstract}

\section{Introduction}

\subsection{``No-go'' theorems}
During the last 70 years the understanding of QM was highly improved
by wide  debate on various ``no--go'' theorems, e.g., von Neumann, 
Kochen--Specker, Bell, see  \cite{VN}, \cite{B}. The latter one really beats all
records on publications, citations, discussions and controversies,
see \cite{FPP}--\cite{FPP3} for recent debates. We emphasize that as any mathematical
theorem a ``no--go'' theorem is based on a number of mathematical
assumptions. And adequacy of a mathematical assumption to physical
reality should be the subject of very careful investigation. For
example, J. Bell criticized strongly some assumptions of the von
Neumann, Jauch-Piron, and Gleason ``no--go'' theorems \cite{B}. Some
assumptions of Bell's theorem were also strongly criticized, see
e.g. \cite{FPP}--\cite{AP}.

\subsection{Probabilistic and quantum contextualities}
In particular, it  was pointed out that the proof of Bell's
inequality is based on the implicit use of a single Kolmogorov
probability space, see Accardi \cite{ACC1}--\cite{ACC4}, Khrennikov \cite{AY1}--\cite{AY2},
Hess and Philipp \cite{HPL2}. We can call such an assumption  {\it
probabilistic non--contextuality.} By {\it probabilistic
contextuality} we understand dependence of probability on
experimental settings. This notion differs essentially from the
conventional notion of quantum contextuality \cite{B}.

We recall that {\it quantum contextuality} is defined as follows:
the result of measurement of an observable $a$ depends on another
measurement on an observable $b,$ although these two observables commute
with each other \cite{B}. It should be emphasized that property of
locality is a special case of quantum non-contexuality.

We now compare conventional quantum contextuality and probabilistic
one.  In some special cases one can obtain probabilistic contextuality from quantum
contextuality. However, probabilistic contextuality need not be
induced by the quantum one:

{\it  The probability distribution can be
dependent on both (commuting) observables even if the result of
measurement of an observable $a$ does not depend on another
measurement on observable $b.$}

\section{Realistic models violating Bell's inequality}
 Many authors studied
different probabilistically contextual models which violate Bell's
inequality. In particular, the efficiency of detectors loophole 
as well as more general the fair sampling loophole, see e.g. 
\cite{Gis}- \cite{Gis4} for these loopholes,  are just special
forms of probabilistic contextuality. In the latter cases different
Kolmogorov spaces correspond to different ensembles of particles
created through selections corresponding to various experimental
settings. By choosing observables $a, b$ and $c,d$ in the EPR-Bohm
framework\footnote{Here $a$ and $b$ as well as $c$ and $d$ 
are orientations of two spatially separated 
polarization beam splitters.} 
we select two different sub-ensembles $\Lambda_{a,b}$ and
$\Lambda_{c,d}.$ Fair sampling assumption means that restrictions of
the probability $P$ (originally defined on the complete  space
$\Lambda$ of hidden variables) onto sub-ensembles $\Lambda_{a,b}$
and $\Lambda_{c,d}$ coincide:
\begin{equation}
\label{CC} P \vert_{\Lambda_{a,b}}= P\vert_{\Lambda_{c,d}}
\end{equation}
This is a special case of probabilistic non-contextuality. And
unfair sampling means that the coincidence condition (\ref{CC}) is
violated for some experimental settings. This is a special case of
probabilistic contextuality.

We remark that in the probabilistic contextual approach one can
derive generalizations of Bell's inequality which are not violated
for quantum covariations \cite{AY1}--\cite{AY2}.

\subsection{Physical origin of probabilistic contextuality}

Thus mathematically everything is clear:  by dropping the assumption
on probabilistic  non--contextuality and assuming that different
experimental settings induce different probability spaces it is
possible  to violate Bell's inequality. But the physical origin of
probabilistic contextuality is a problem of huge complexity. In all
conventional models probabilistic contextuality is induced either by
quantum contextuality or by losses of particles\footnote{E.g.,
efficiency of detectors, fair sampling, and time--window loopholes
\cite{Gis}--\cite{Gis4}, induce losses of particles: a part of the original ensemble
should disappear. We agree that losses of particles is the important
problem. However, we do not think that this is the essence of Bell's
argument. We agree with experimenters that such losses of particles
can be considered merely as a technological problem. One of the
authors would like to thank Alain Aspect and Gregor Weihs for
discussions on this problem during V\"axj\"o conferences.} On the
other hand, we do not know any natural physical explanation of
quantum contextuality, besides nonlocality.

\subsection{Chameleon effect}

However, there  exists a model in that probabilistic contextuality
(i.e., dependence of probabilities on experimental settings) can be
produced without losses of particles. Moreover, in that model
probabilistic contextuality is not a consequence of the quantum
contextuality and hence the model is {\it local.}

This is {\it the chameleon model} which is described in detail  
\cite{ACC1}--\cite{ACC4}.  In
these papers  Bell's definition of realism  was  criticized and
there was proposed a new approach to realistic models, namely, {\it
adaptive realism.} In the chameleon model one could not identify
results of measurements with ontic variables (i.e., preexisting
before measurement). Suppose a particle has some property, say spin.
At the ontic level spin is characterized by some parameter $\sigma$.
{\it Can one assert that precisely this parameter is obtained as the
result of a spin--measurement?} Definitely not! Any measurement is a
complicated process of interaction of a microscopic system with a
measurement device. Finally we  cannot say that we obtain the ontic
parameter $\sigma,$ but only the observed spin, say $S$. We
emphasize that QM as about $S$ and not about $\sigma$ (as N. Bohr
pointed out in many occasions QM is not about reality as it is, but
about the results of measurements).

How does the result of measurement $S$ arise?  This is the result of
{\it dynamical process} of interaction of a system and a measurement
device. In such an approach there as nothing against realism.
However, this is the adaptive (or chameleon) realism (which is not
at all realism of balls having once and for ever determined color).

The chameleon effect simply states that,  since dynamics  is
determined by the variable subjected to measurement, we  obtain
probability distributions depending on experimental settings. Thus
the chameleon approach implies probabilistic contextuality, hence,
the possibility of violation of Bell's inequality. Nevertheless,
dynamics of measurements can be completely local. Let $a$ and $b$ be
two quantum observables represented by commuting operators. Then
there are two different dynamical systems corresponding to the $a$
and $b$-measurements, respectively. In general, they do not depend
on each other. Therefore the chameleon effect induces probabilistic
contextuality, but not at all quantum contextuality.

Finally, we remark that we  question neither Bell's theorem as a
mathematical result nor experimental violation of Bell's
inequality. We question the adequacy of Bell's realistic model
(which he used to confront classical and quantum physics) to the
physical situation. We show that by rejecting two basic implicit
assumptions in Bell's definition of a realistic model, namely

\medskip

a) non-adaptive realism of observables;

b) the range coincidence hypothesis,

\medskip

we can construct a model with hidden variables which reproduces
precisely the EPR--Bohm correlations.

\section{Forward and backward  Kolmogorov equations}

Our  further considerations generalize the well known dynamical
scheme for statistical states and variables associated with the
diffusion process. Therefore we recall the standard scheme. Let $x(s)$ be a diffusion process. To simplify considerations, we 
consider at the beginning the state space $X={\bf R},$ the real line. We set
$$p(s,x,t,y)=P(x(t)=y|x(s)=x)$$
We consider the probability measure (statistical state describing an
ensemble of particles)
\begin{equation}
p(s,t,y)=\int p(s,x,t,y)p_0(x)dx, \label{z1}
\end{equation}
where $p_0(x)$ is the density of the initial probability distribution on the state 
space.
This probability satisfies to the forward Kolmogorov equation:
\begin{equation}
{\partial p(s,t,y)\over\partial t}\,=L(p(s,t,y))\ ,\label{k1}
\end{equation}
where the generator of diffusion is given by
\begin{equation}
L(p)(t,y)={1\over2}\,{\partial^2\over\partial
y^2}\,[\sigma^2(t,y)p(t,y)]-{\partial\over\partial
y}\,[a(t,y)p(t,y)]\ .\label{k2}
\end{equation}
Here $a(t,y)$  and $\sigma(t,y)$ are the drift and diffusion
coefficients, respectively. We note that in physics (\ref{k1}) is
known as the Fokker--Planck equation. The evolution equation
(\ref{k1}) is completed by the initial condition:
\begin{equation}
\lim_{t\downarrow s}p(s,t,y)=p_0(y)\label{k3}
\end{equation}
Let us now consider the corresponding dynamics of functions. We set
\begin{equation}
f(s,\tau,x)=\int g(y)p(s,x,\tau,y)dy\label{z2}
\end{equation}
Then this function satisfies to the backward Kolmogorov equation:
\begin{equation}
{\partial f\over\partial s}\,(s,\tau,x)=W(f(s,\tau,x))\label{z3}
\end{equation}
where the operator $W$ which is conjugate to the generator $L$ is given by
\begin{equation}
W(f)(s,x)=-{1\over2}\,\sigma^2(s,x){\partial^2 f(s,x)\over\partial
x^2}\,-a(s,x){\partial\over
\partial x} f(s,x)\ .\label{z4}
\end{equation}
The evolution equation (\ref{z3}) is completed not by initial condition, but by the ``final condition'':
$$\lim_{s\uparrow \tau}f(s,\tau,x)=g(x)$$
We emphasize this crucial difference between the equations for
statistical states (probabilities) and physical variables (functions
on the configuration space). The former is {\it a forward equation}
and the latter is {\it a backward equation.} By knowing a
probability distribution $p_0(y)$ at the initial instance of time
$s=t_0$ we can find it at any $t\geq t_0$: $p(t_0,t,y)$. By knowing
a physical variable $g(y)$ at the end of evolution $t=\tau$ we can
reconstruct it at the initial instance of time $t_0:$
$f(t_0,\tau,x)$.

We remark that
$$
\int f(t_0,\tau,x)p_0(x)dx=\int\left(\int
g(y)p(t_0,x,\tau,y)dy\right)p_0(x)dx
$$
$$
=\int g(y) \left(\int p(t_0,x,\tau,y)p_0(x)dx\right) dy=\int
g(y)p(t_0,\tau,y)dy.
$$
Since in our further considerations we will not always be able to
operate always with densities, we consider just probability
measures: $p_0(dy)$, $p(s,t,dy)$ and so on. We rewrite the forward
and backward Kolmogorov equations in the compact form:
\begin{equation}
{\partial p(t_0,t)\over\partial t}\,=L(p(t_0,t))\ ,\quad p(t_0,t_0)=p_0\ ;\label{c1}
\end{equation}
\begin{equation}
{\partial f\over\partial s}\,(s,\tau)=W(f(s,\tau))\ ,\quad
f(\tau,\tau)=g\label{c2}
\end{equation}
We have the following conjugation condition:
\begin{equation}
\int f(t_0,\tau,x)p_0(dx)=\int g(y)p(t_0,\tau,dy)\label{c3}
\end{equation}
or
\begin{equation}
\int f(t_0,\tau,x)p(t_0,t_0,dy)=\int
f(\tau,\tau,x)p(t_0,\tau,dx)\label{c4}
\end{equation}
We remark that  only one quantity, either 
a probability measure or a function, is known in each side of this equality.

The Cauchy problem (\ref{c1}) induces the dynamical system $V_{t_0,t}$ in the space of probability measures:
\begin{equation}
p(t_0,t)=V_{t_0,t}(p_0)\ ,\label{c5}
\end{equation}
and the (backward) Cauchy problem (\ref{c2}) induces the dynamical
system $U_{s,\tau}$ in the space of functions:
\begin{equation}
f(s,\tau)=U_{s,\tau}(g)\label{c6}
\end{equation}
These dynamical systems are conjugate:
\begin{equation}
\int U_{t_0,\tau}(g)(x)p_0(dx)=\int
g(x)V_{t_0,\tau}(p_0)(dx)\label{c7}
\end{equation}

\section{Classical statistical model with the chameleon effect}

Denote by $\Lambda$ the state space of physical systems under
consideration. We also consider statistical states describing
ensembles of systems. They are represented by probability measures
on $\Lambda$. Physical variables are represented by functions
$f:\Lambda\to\Bbb R$. The average  of a variable $f$ with respect to
a statistical state $p$ is given by
\begin{equation}
\langle f\rangle_p=\int_\Lambda f(\lambda)p(d \lambda)\label{A2}
\end{equation}
Dynamics of a statistical state is given by a dynamical system
$V_{t_0,t}$ in the space of probability measures. Dynamics of a 
physical variable is given by a dynamical system
$U_{s,\tau}$ in the space of functions. 

We no longer
assume that these dynamics are generated by a diffusion (not even a
Markov process). The $V_{t_0,t}$ and $U_{s,\tau}$ are two general
dynamics. The only condition coupling them is the conjugation
condition (\ref{c7}).

We emphasize that $V_{t_0,t}$ is the forward dynamics: by knowing
the initial statistical state, $p_0$, we can find it at any instant
of time $t$: $p(t_0,t)=V_{t_0,t}(p_0)$. In contrast, $U_{s,\tau}$ is
the backward dynamics: by knowing the final physical variable
$f_\tau(x)=g(x)$, we can reconstruct it for the $t=t_0$:
$f(t_0,\tau,x)=U_{t_0,\tau}(g)(x)$. This was the well known story.
The chameleon story starts when one wants to describe processes of
measurements.

Suppose that we would like to present classical statistical  (but
dynamical!) description of the process of measurement of an
observable $a$. Here $a$ is just a label to denote a class of
measurement devices. In QM we use self--adjoint operators as such
labels.

In the chameleon model of measurement the basic assumption is  that
dynamics $V$ and $U$ depend on the observable $a:$
\begin{equation}
\label{c5Z}
V_{t_0,t}\equiv V^a_{t_0,t}\ ,\quad U_{s,\tau}\equiv U^a_{s,\tau}.
\end{equation}
This is a very natural assumption: {\it any measurement device
changes dynamics.} Suppose that initially there was prepared an
ensemble of systems with the probability distribution
$p_0(\lambda)$. Then in the process of the $a$--measurement
$p_0(\lambda)$ evolves according to the dynamics $V^a$.

We assume that  the process of measurement takes the finite interval
of time $\tau$. Thus at that moment the probability distribution
becomes $p_\tau(\lambda)$ (which is, of course, depends on $a).$

\medskip

The physical variable $f^a_t(\lambda)$ evolves according to the
dynamics $U^a$. We do not know the initial (ontic) physical variable
$f^a_{t_0}(\lambda)$. This is a hidden physical variable -- an ontic
property of systems before the $a$--measurement starts. In our model
a particle has the ontic position, momentum, spin and so on. But it
would be very naive to expect (as J. Bell did) to measure directly
$f^a_{t_0}(\lambda)$. We measure the result of evolution, namely,
$f^a_\tau(\lambda)$. The latter variables are the results of
measurements. QM is, in fact, about such variables. But, in contrast
to the chameleon model, QM does not permit the functional
representation of observables.

We repeat again that dynamics for variables is a backward dynamics.
Such a mathematical description is totally adequate to the physical
experimental situation. We do not know the initial variable
$f^a_{t_0}(\lambda)$, but only the final (observed) variables
$f^a_\tau(\lambda).$

We can reconstruct $f^a_{t_0}(\lambda)$  from the observed quantity
$f^a_\tau(\lambda)$. But we never know $f^a_{t_0}(\lambda)$ from the
very beginning. Therefore we are not able to construct
$f^a_\tau(\lambda)$ and hence predict the result of measurement.

We have two types of averages:

(CL) The ontic (``classical'') averages are given by
\begin{equation}
\langle f^a\rangle_{CL}\equiv\langle
f^a_{t_0}\rangle_{p_0}=\int_\Lambda f^a_{t_0}(\lambda)p_0
(d\lambda)\ ;\label{A12}
\end{equation}
$(OB)$ The observational averages (in particular, the quantum ones) are given by
\begin{equation}
\langle f^a\rangle_{OB}\equiv\langle f^a_\tau\rangle_{p^a_\tau}=\int
f^a_\tau(\lambda)p^a_\tau(d\lambda)\ . \label{A13}
\end{equation}
As a consequence of the conjugation condition (\ref{c3}), these
averages coincide:
\begin{equation}
\langle f^a\rangle_{CL}=\langle f^a\rangle_{OB}\label{c3A}
\end{equation}
Thus one can either consider the average with respect to the initial
probability distribution: $\langle f^a\rangle_{CL}$, but the $f^a$
be the ontic variable and not the observed one, or the average of
the observed physical variable, $\langle f^a\rangle_{OB}$, but in
this case the initial probability distribution $p_0$ could not be
used. In the latter case one should consider the probability we
assure $p^a_\tau$  that depends on $a$.

In the special case of quantum measurements the (OB) gives the quantum average and the average
(CL) can be called prequantum. In the model under consideration we assume that 
the quantum and prequantum averages coincide. Recently there was proposed a model, Prequantum Classical 
Statistical Field Theory, producing a prequantum average which coincides with the quantum 
one only approximately, see \cite{KH1}--\cite{KH4}.

Finally we remark that if $f^a_\tau$ takes, e.g., the values
$\{\pm1\}$, then there will  be no reasons to assume that $f^a_0$
takes the same values.

\subsection{The range of values coincidence hypothesis}

Recently it was paid attention, see  \cite{KH1}--\cite{KH4}, to another problem in Bell's
definition of realism \cite{B}. This is the {\it range of values
coincidence problem:}

\medskip

{\it A priori there are no reasons to assume that the range of
values of an ontic physical variable (say ontic spin $\sigma)$
should coincide with the range of values of the corresponding
observables (say measured spin $S).$}

\medskip

As was already pointed out, the process of measurement is the
process of interaction of a microscopic system and a measurement
device. Therefore it is not surprising that the $\sigma$ can be
transformed into a different value $S$. In fact, by its very
definition $\sigma$ is unobservable in principle.

Denote by $\lambda$ the state of a system, the ``hidden variable''.
Both $\sigma$ and $S$ are functions of
$\lambda:\sigma=\sigma(\lambda)$, $S=S(\lambda)$. But there are no
reasons to assume that
\begin{equation}
\hbox{Range }\sigma=\hbox{ Range } S\ .\label{A}
\end{equation}
Thus one should sharply distinguish  ontic and observed variables.
The condition that the observed spin $S=\pm1$ does not imply that
the ontic spin $\sigma$ (which is in principle unobservable) also
takes values $\pm1.$

\subsection{Spectral postulate}

We point out that we do not want to drop the standard {\it spectral
postulate} of QM. By this postulate the range of vales of a quantum
observable coincides with the spectral set of the corresponding
self-adjoint operator. This postulate was confirmed by all quantum
experiments and it could not be questioned. In our approach the
range of values of say  the observed spin $S$ coincides with the
spectral set of the corresponding quantum operator. We simply remark
that there is no reasons to expect that the range of values of say
the ontic spin $\sigma$ should coincide with this spectral set.

\subsection{Classical reproduction of the EPR-Bohm correlations}

In fact, we need not to consider a new classical adaptive (chameleon) model 
which would give us the EPR-Bohm correlations. By taking into account the analysis 
of measurement process which was performed in the present paper (and especially 
the evident possibility of violation of the range of values coincidence hypothesis)
we can now use the well known model of  Accardi and Regoli \cite{ACC5}. 

{\bf Conclusion:} {\it The common conclusion that Bell's arguments imply 
incompatibility of local realism and the quantum formalism is based on a rather 
naive understanding of coupling between ontic reality (i.e., reality as it is when nobody 
make measurements) and the observational reality. By considering the adaptive measurement 
framework (based on the chameleon effect) we showed that in fact local realism can 
peacefully coexist with the quantum formalism.}

\end{document}